# Analysis of the Performance of Dynamic Multicast Routing Algorithms

Jaihyung Cho, James Breen
School of Computer Science and Software Engineering
Monash University
{jaihyung,jwb}@dgs.monash.edu.au

## Abstract

*In this paper, three new dynamic, low delay multicast routing algorithms based on the greedy tree technique are proposed; Source Optimised Tree, Topology Based Tree and Minimum Diameter Tree. A simulation analysis is presented showing various performance aspects of the algorithms, in which a comparison is made with the greedy and core based tree techniques. The effects of the tree source location on dynamic membership change are also examined. The simulations demonstrate that the Source Optimised Tree algorithm achieves a significant improvement in terms of delay and link usage when compared to the Core Based Tree, and greedy algorithm.*

## 1. Introduction

In recent years, various static [4][17][18] or dynamic [1-3][6][19-21] multicast routing algorithms have been proposed. In the situation where the group membership is continually changing and real time communication is required, complex algorithms are not well adapted to multicast communications. The overhead required in rebuilding the tree whenever new join or leave requests arrive is considered too expensive. The simple approaches, such as attaching a new node at a suitable location on an existing tree, are more feasible because such methods require only partial knowledge of the network configuration and less computational overhead, and the routing decision can be made by each joining node. One category of these algorithms [1][2][4][6][19-21] are known as the shortest path based algorithm because they adopt shortest path tree for multicast data distribution.

When there are many sources in a group communication, a multicast route by the shortest path algorithm can be provided by multiples source specific trees [19][21] or a shared type tree [6][20]. The shared type shortest path tree is simple to implement and requires less state information than the former. However the overall performance is sensitive to the root location [7] and the delay performance is not good when the source is not located in the root node. Later in this paper, we will present simulation results showing the extent of the performance variation associated with the core location.

The other type of simple algorithm is the core-less, branch attachment method, widely known as the Greedy algorithm (GRD) [1][2][14]. GRD was originally known as the Nearest Node [13] algorithm or Nearest Insertion [3] algorithm, and was developed as a heuristic to the minimum cost Steiner tree. This method finds the closest attachment point in tree and connects the new node to that point via the shortest path. This heuristic requires a relatively low computational overhead, scales well and achieves low cost multicast tree. This method, however, has several drawbacks:

(a) it does not constrain the maximum path length, and hence the delay performance can be worse than other trees, such as the shared shortest path tree.

(b) the performance fluctuates severely as members are added and deleted, as noted by Waxman [1] and Dore [5], and confirmed by our simulation results.

(c) the joining node may need to know the entire multicast group configuration information in order to resolve the best attachment point of the branch path.

There have been several proposals to optimize the delay performance based on greedy algorithm. [8][9] However these algorithms perform repetitive computation in order to constraint delay within a bound, hence they require entire network knowledge and it is hard to apply to distributed routing.

In this paper, three new algorithms are proposed which improve the delay performance of GRD. They are the Source Optimised Tree (SOPT); Topology based Optimisation Tree (TOPT), which draws upon an existing unicast distance table in order to reduce the overhead of nodal weight calculation; and the Minimum Diameter Tree (MDT), which constraints the maximum length of the tree. The various performance aspects of these algorithms have been compared against GRD and CBT, and we propose the Source Optimised tree as an efficient dynamic-shared multicast tree in terms of delay and bandwidth consumption.

## 2. The Greedy tree based algorithms

### 2.1 Earlier Algorithms

The mechanism of the GRD is to find the closest attachment point from the new participating node to the existing tree and to connect to it via the shortest path. This can be described formally as following:

(A1) given a tree sub-graph $\mathbf{T}_{n-1}$ and a member
node $\boldsymbol{v} \notin \mathbf{T}_{n-1}$,
find the best attachment point $\boldsymbol{a} \in \mathbf{T}_{n-1}$,
such that the new sub-graph
$\mathbf{T}_n = \mathbf{T}_{n-1} + \mathbf{p}(\boldsymbol{a}\text{-}\boldsymbol{v})$, (where $\mathbf{p}(\boldsymbol{a}\text{-}\boldsymbol{v})$ is the
shortest path) is close to the optimum.

This is a relaxed algorithmic definition of GRD because the statement does not specify criteria for the 'best attachment point' and the 'optimal' state of the tree. Depending on the definition of the optimal tree and the attachment point selection algorithm, different types of trees can be obtained. In this paper, we have defined the optimality of the multicast tree to include the minimum usage of the bandwidth, lowest delay, and stability during changes to the multicast group membership. Details of the evaluation methods of these performance factors will be given in section 3.

In the definition of (A1), the branch path $\mathbf{p}(\boldsymbol{a}\text{-}\boldsymbol{v})$ can be chosen via the shortest delay route, the maximum available bandwidth route or the minimum cost route, according to the required QoS. The simple and practical method is to use the shortest path according to the same routing metric used by unicast routing, such as hop count or delay.

Provided that the shortest path is used for grafting the branch path, different shapes of trees can be built, with different performance characteristics, depending upon the attachment point selection algorithm. When the attachment point $\boldsymbol{a}$ is chosen to minimise the length of the path to the root node, a shortest path tree is established. When the closest attachment node in the tree is selected, the GRD tree is established. The cost performance of the dynamic greedy tree is better than the shared shortest path tree [1][2], and has an upper bound of ½ log n times that of the optimal Steiner tree, where n is the number of members [14]. However the delay performance of GRD is worse than shared shortest path trees [3]. The GRD mechanism only considers the length of the new branch path connecting the joining member, not the extended length of the tree after grafting. Therefore consecutive join operations at the end of the leaf nodes may result in trees with extended maximum length.

Waxman has suggested a weighted greedy tree [2] as an improvement of GRD. In his proposal, an owner node $\boldsymbol{o}$ of a group is selected and the node remains in the tree throughout the session. A weighted greedy tree is established by choosing the attachment node $\boldsymbol{a}$, which minimises the weight function:

(A2) $\mathbf{W}(a,v) = (\mathbf{1}-\omega)*\mathbf{d}(v\text{-}a) + \omega*\mathbf{d}(a\text{-}o)$,
where $0 <= \omega <= 0.5$

The $\omega$ factor is a weight constant which controls the ratio of the branch length and the distance from the attachment point $\boldsymbol{a}$ to the owner node $\boldsymbol{o}$. When $\omega$ decreases towards zero, the tree shape approaches a GRD tree and when $\omega$ increases toward 0.5, the tree approaches the shortest path tree rooted at the owner node $\boldsymbol{o}$. Therefore, the weighted greedy tree has a performance characteristic which lies between the GRD and the shared shortest path tree (or CBT). From an empirical study, Waxman suggested 0.3 as an optimal value for $\omega$.

### 2.2 Proposed New Algorithms

Waxman's method is viewed as an improvement of CBT, although it was proposed earlier. Indeed, the weighted greedy tree has the same defects as CBT. For example;

(a) the performance of the weighted greedy algorithm is affected by the location of the owner node $\boldsymbol{o}$ and it leaves open the question of how to choose the owner node $\boldsymbol{o}$ in order to achieve the best performance.

(b) while Waxman's weighted greedy algorithm performs well when the owner node $\boldsymbol{o}$ becomes the only source node, the performance can degrade as multiple sources share the tree.

(c) the owner node $\boldsymbol{o}$ can be the centre of a region of congestion and the failure of the owner node is critical.

Therefore the weighted greedy algorithm does not overcome the problems related to the location of the core node. However, Waxman's approach with the weighted greedy function provides a mechanism for improving the performance of GRD.

The first algorithm is the Source Optimised Tree (SOPT) algorithm, which minimises the delay from the most distant source to the joining node. The algorithm is described as following:

(A3) Given a tree sub-graph $\mathbf{T}$ and a joining
node $\boldsymbol{v} \notin \mathbf{T}$, let $\mathbf{S}$ be the
set of all sources sharing the tree $\mathbf{T}$.
A source optimised tree is built
by grafting the shortest path $\mathbf{p}(\boldsymbol{v}\text{-}\boldsymbol{a})$ on the
node $\boldsymbol{a} \in \mathbf{T}$, such that
the following function is minimised.

$\mathbf{W}(a,v) = \mathbf{d}(v\text{-}a) + \omega * \mathbf{Max}\{\mathbf{d}(s\text{-}a) \mid \mathbf{for\ all}\ (s \in \mathbf{S}),\}$, where $\omega > 0$

In above function, the weight of each of the candidate attachment points of the tree **T** is computed by measuring the maximum distance from the node to all sources via the tree **T**, e.g, by monitoring the packet delay from all sources. The joining node is connected via the path that minimises the length (or delay) of the branch path plus the weighted maximum delay of the attachment point, thus minimising the maximum data reception delay.

In function (A3), the weight constant $\omega$ is assigned only to the second term and must have a value greater than zero. When $\omega$ decreases to zero, the tree shape approaches the GRD tree. For one source and $\omega$ approaching unity, SOPT produces a shortest path tree rooted at the source node. Therefore the performance of the SOPT algorithm lies between that of the GRD and the CBT in the presence of single source. However, significant performance improvements can be seen when multiple sources share the SOPT tree. (See. section 4)

The second algorithm is the Topology based Optimisation Tree (TOPT), which utilises the distance information of the underlying unicast system, and thereby reduces the computational overhead. In this algorithm, the multicast-capable nodes in the network compute the average distance to all known destinations in advance, using the distance information in the unicast routing table. When a new member requests to join the existing tree, the algorithm finds the attachment node which makes the length of the branch path plus the weighted average distance a minimum. This algorithm applies concepts from the optimum transportation network design problem [15][16], in which the node with a small average distance to all other nodes can serve as a low delay distribution centre in the network. Therefore, when the candidate attachment points of a tree are evaluated for grafting a new branch, the node located near the topological centre of the network is preferred. The TOPT algorithm is described as following;

(A4) Given a tree sub-graph **T** and a member
node $v \notin$ T, let **N** be the
set of all nodes in the network.
A Topology based Optimisation Tree is
built by grafting the shortest path $\mathbf{p}(v\text{-}a)$
on the node $a \in \mathbf{T}$, such that the
following weight function is minimised;

$$\mathbf{W}(a,v) = \mathbf{d}(v\text{-}a) + 2\omega * \left( \sum_{n \in \mathrm{N}} \mathbf{d}(a\text{-}n) \right) / |\mathbf{N}| \quad \text{where } \omega > 0$$

The second term in the function (A4) is the calculation of the weighted average distance of the node $a$. The average distance is evaluated each time the unicast routing tables are updated. Since the average distance is associated with the node and is independent of the formation of the multicast group, the same average distance value can be applied for different groups served by a node, unless the network configuration changes significantly.

The shape of the TOPT tree is similar to Wall's centre based tree [4] when the population of the group is evenly distributed in the network. (See Figure 4). However the centre based tree becomes inefficient when many of the members are concentrated in a locality and located some distance from the centre node, because every group member has to include the centre node as a root. In contrast, TOPT constructs an efficient tree which serves only the locally concentrated nodes.

The third algorithm is the Minimum Diameter Tree (MDT) algorithm, which constructs the tree by grafting the new branch path at the point where the distance from the joining node to the most distant leaf node is a minimum. Thus MDT attempts to constrain the maximum length (or diameter) of the tree. The algorithm is described as:

(A5) Given a tree sub-graph **T** and a joining
node $v \notin$ T, let **E** be the set of
all leaf nodes of the tree **T**. A minimum
diameter tree is built by grafting
the shortest path $\mathbf{p}(v\text{-}a)$ on the point
$a \in \mathbf{T}$, such that it minimises
the weight function;

$$\mathbf{W}(a,v) = \mathbf{d}(v\text{-}a) + \omega * \mathbf{Max}\{\mathbf{d}(e\text{-}a) \mid \text{for all } (e \in \mathbf{E})\}, \text{ where } \omega > 0$$

The weight function of (A5) is similar to that in (A3) except that the distance to the most distant leaf nodes is calculated instead of the distance to the source nodes. The diameter of a tree is defined as the longest distance between any two leaf-nodes of the tree. According to the lemma proven by Wall [4], the most distant node from any node in the tree is at one end of the diameter. This algorithm, therefore, requires a diameter update mechanism. One possible mechanism is that all leaf nodes report the distance information to other leaf nodes periodically. The intermediate nodes can filter the packets except the one with the longest distance value. As a result, only the two packets issued from both ends of the diameter, and containing the longest distance information, completely propagate through the tree and update the node weight.

The detailed mechanism for the implementation of the three algorithms is beyond the scope of the paper.

## 3. The Network Model and Evaluation Criteria

The simulations reported in this paper were performed using a randomly generated network model proposed by Waxman. For each simulated network, 200 network nodes, representing the network routers, are scattered on a 1000 by 1000 grid.

Each simulation was carried out on networks of average degree 2.5, 3, 3.5, 4, 4.5 and 5, and the effect of the network complexity was observed. For modelling the activity of applications, two scenarios are defined;

(a) a static source model in which the source locations are fixed throughout the session, however the receiving nodes are free to change membership. The simulation covered cases with 1, 2, 3 and 5 source nodes, with the source node locations being varied in each run.

(b) a dynamic source model where the 10%, 50%, and 90% of the member nodes act as a source and receiver.

The purpose of this simulation is to observe the performance efficiency of different tree models as the tree size is expanding or contracting. Therefore the group size is varied at random between 5 to 90, and the performance is measured when the group size reaches the target size of 10, 20,40 and 80.

Each of the 5 algorithms; CBT, GRD, SOPT, TOPT and MDT, was applied to establishing a tree for the same multicast group, and the average and maximum delay, and link usage performance was evaluated. For the CBT algorithm, the arbitrarily selected initial source node becomes the core node. Figures 1~5 show examples of the trees generated by the 5 algorithms in the case of 4 fixed sources and 50 receivers.

In this simulation, the length of the link represents the delay of the link. Therefore the shortest path equates to the shortest delay route. The average delay of a multicast tree is calculated by the mean of the transmission delay from all sources to all receivers. The maximum delay is the longest delay observed from a source to receiver via the shared tree.

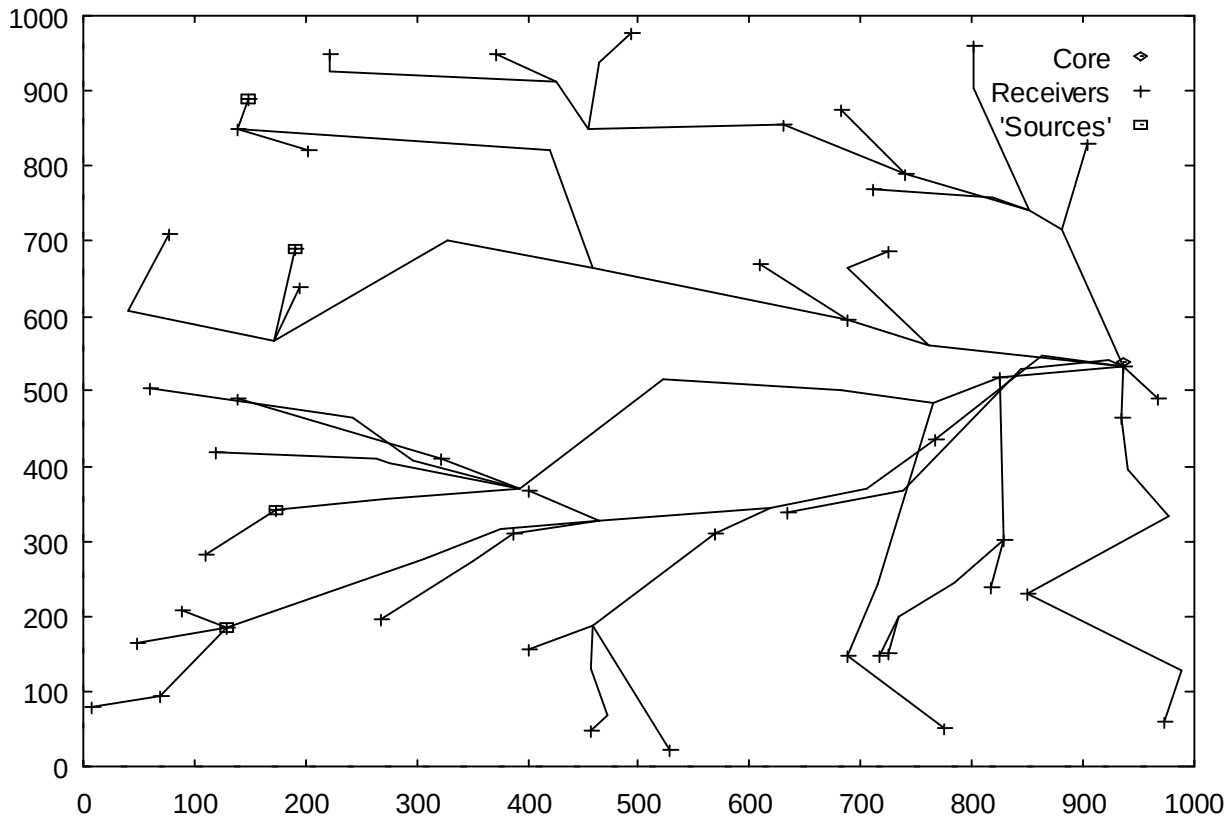

**Figure 1 CBT tree (4 srcs, 50 rcvers)**

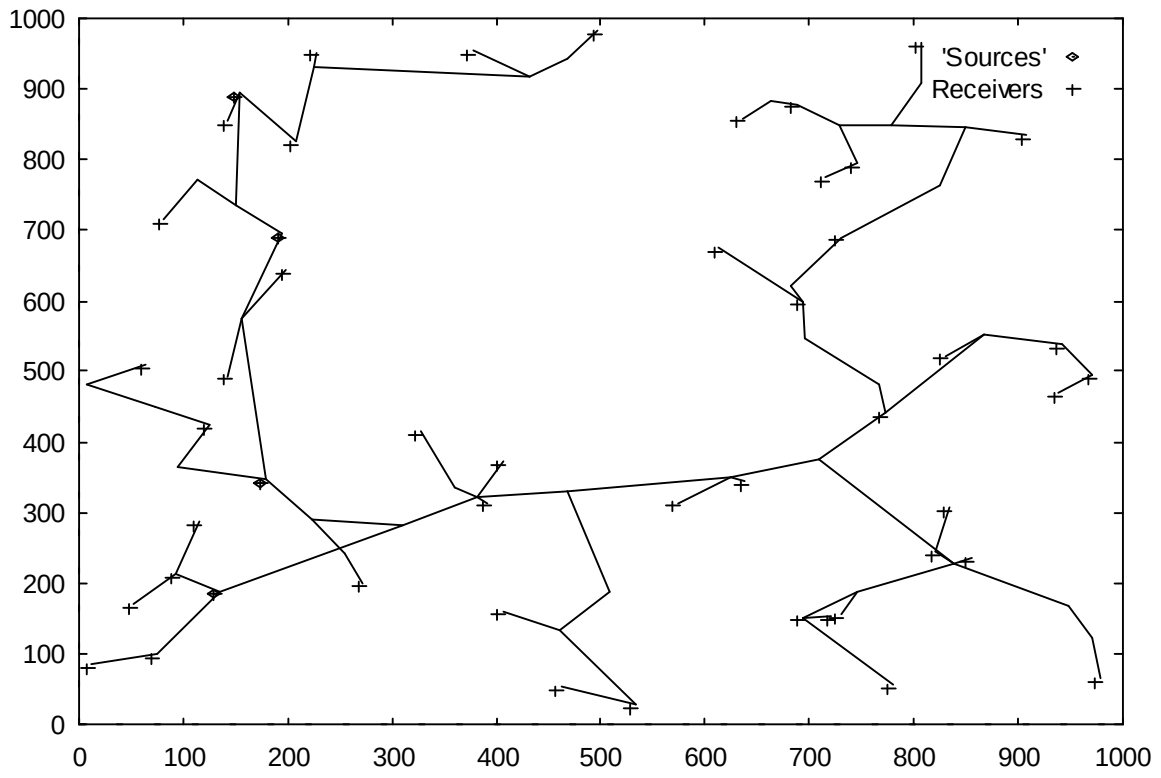

**Figure 2 GRD tree (4 srces, 50 rcvers)**

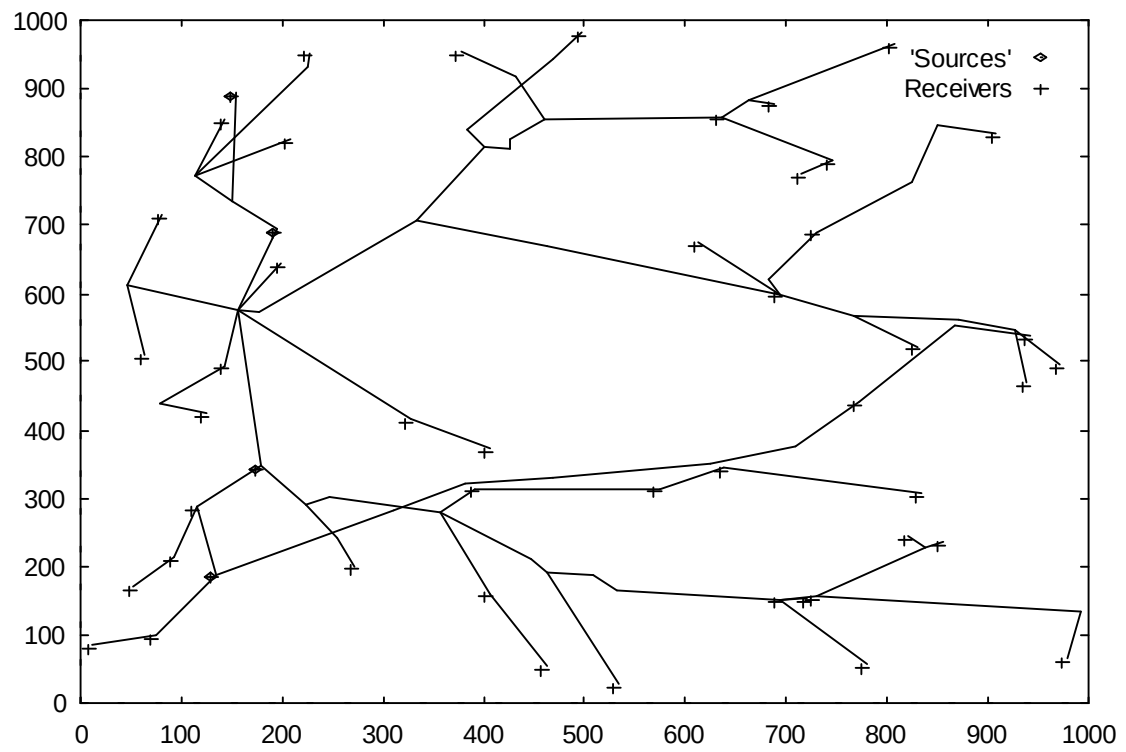

**Figure 3 SOPT tree (4 srcs, 50 rcvers)**

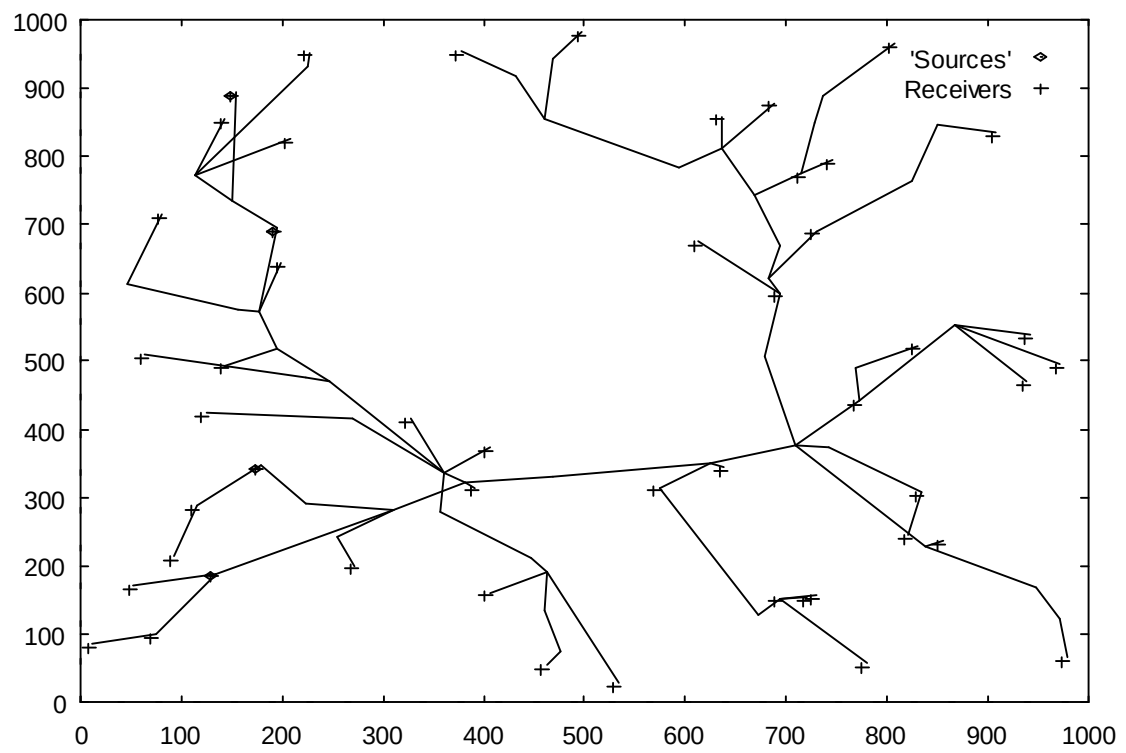

**Figure 4 TOPT tree (4 srcs, 50 rcvers)**

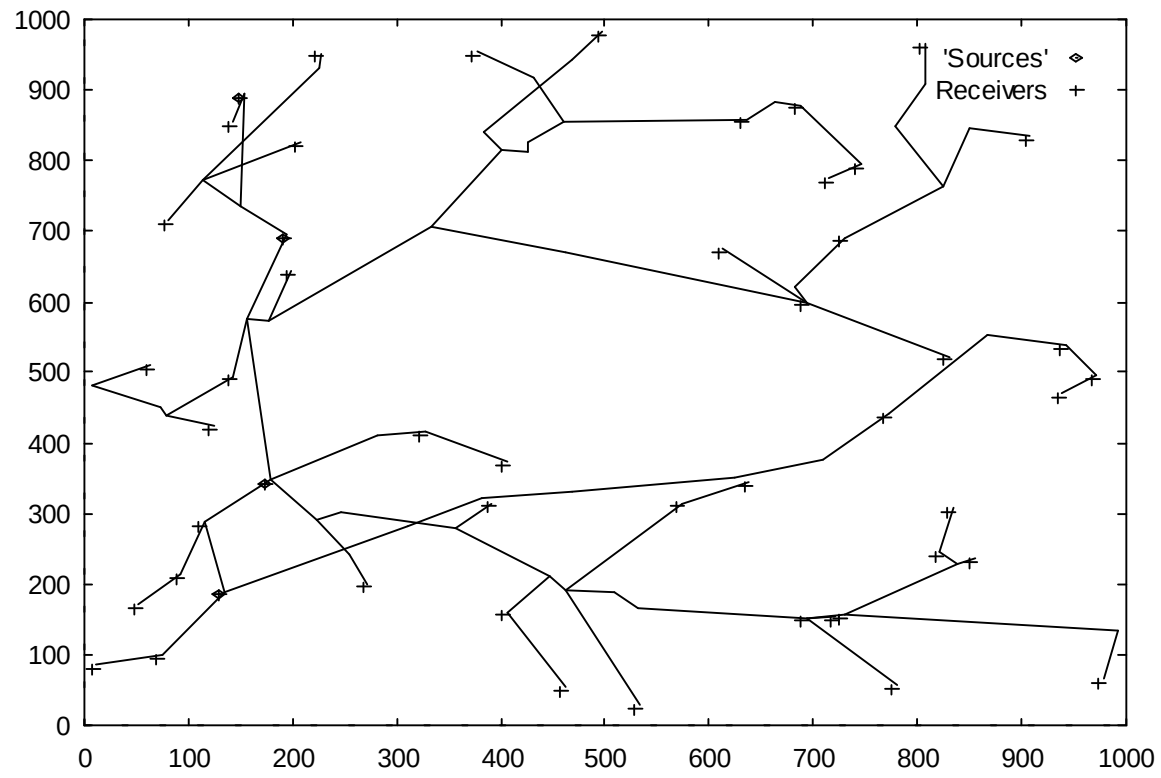

**Figure 5 MDT tree (4 srces, 50 rcvers)**

The link usage, which is estimated by the number of links, is considered. This measure also can be applied to evaluating the bandwidth consumption by the multicast tree. The bandwidth consumed at each link is equal to or less then the output rate of the source. Therefore, an estimate of the total bandwidth usage of a route is given by the product of number of links and the source output transmission rate.

# 4. Simulation Results

## 4.1 Selection of optimal weightings

The efficiency of SOPT, TOPT and MDT is affected by the $\omega$ value (weight). Figure 6 shows an example of the efficiency of these trees compared with CBT as the $\omega$ value increases, using 3 sources and variable number of receivers in a degree 3 network. Repeated simulation using different source numbers and network degrees showed a similar pattern, with little variation in the location of the optimal $\omega$ value, and only differing in the curvature and height. It is clear from inspection of the graphs that there is an optimal $\omega$ value for each algorithm. Table 1 is the summary of the optimal $\omega$ values extracted from over 20,000 runs. As a result, we propose 0.6, 0.8 and 0.4 as appropriate optimal $\omega$ values for using in the SOPT, TOPT and MDT respectively. The simulation results, which follow in this paper, are carried out using these optimal values.

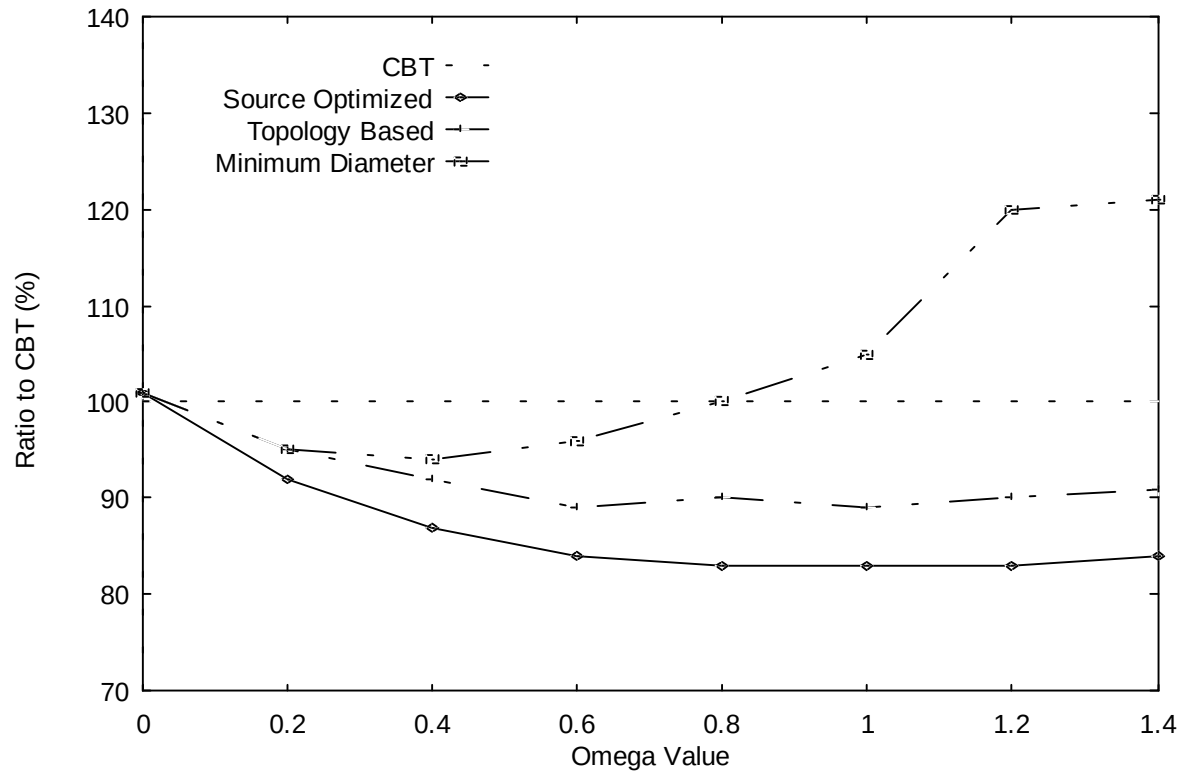

**Figure 6 Maximum path delay with Varying $\omega$ (3 srces in degree 3 network)**

| | SOPT | TOPT | MDT |
|---|---|---|---|
| Optimum for Average Delays | 0.6 | 0.8 | 0.2 |
| Optimum for Maximum Delays | 0.8 | 0.8 | 0.4 |
| Proposed $\omega$ values | 0.6 | 0.8 | 0.4 |

**Table 1 summary of the optimal $\omega$ values**

## 4.2 Delay variation stability

An important issue in the stability of multicast routing algorithms is whether their performance is stable when they are subject to repeated join and departure operations. For this reason, each algorithm was subjected to simulation runs of 200,000 membership changes. Figures 7~10 show examples of the maximum delay variation during these runs, and depict the multicast tree performance for groups of 40 receivers and 5 fixed sources. Each simulation was repeated 200 times and the worst-case and best-case graphs are presented to contrast the extent of the performance difference. The five algorithms exhibited different patterns as multiple sources share the tree.

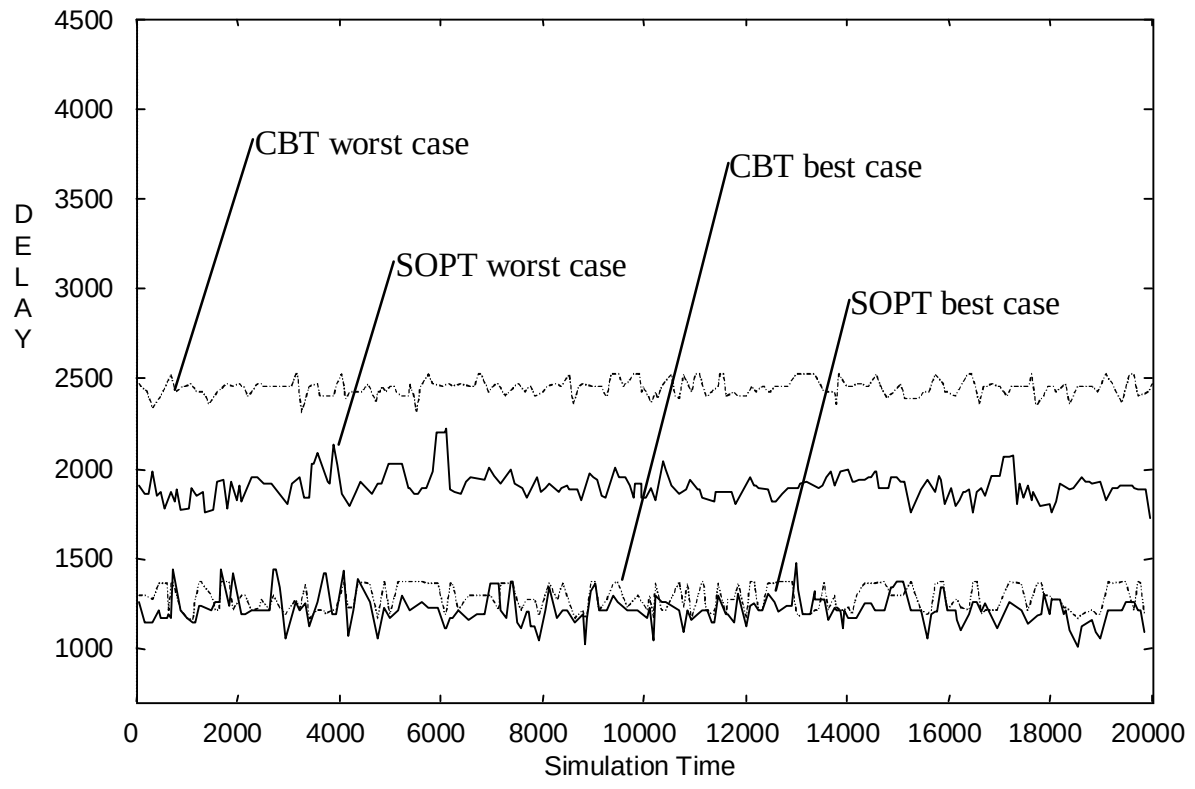

**Figure 7 Maximum delay pattern of CBT and SOPT (5 sources, degree 5)**

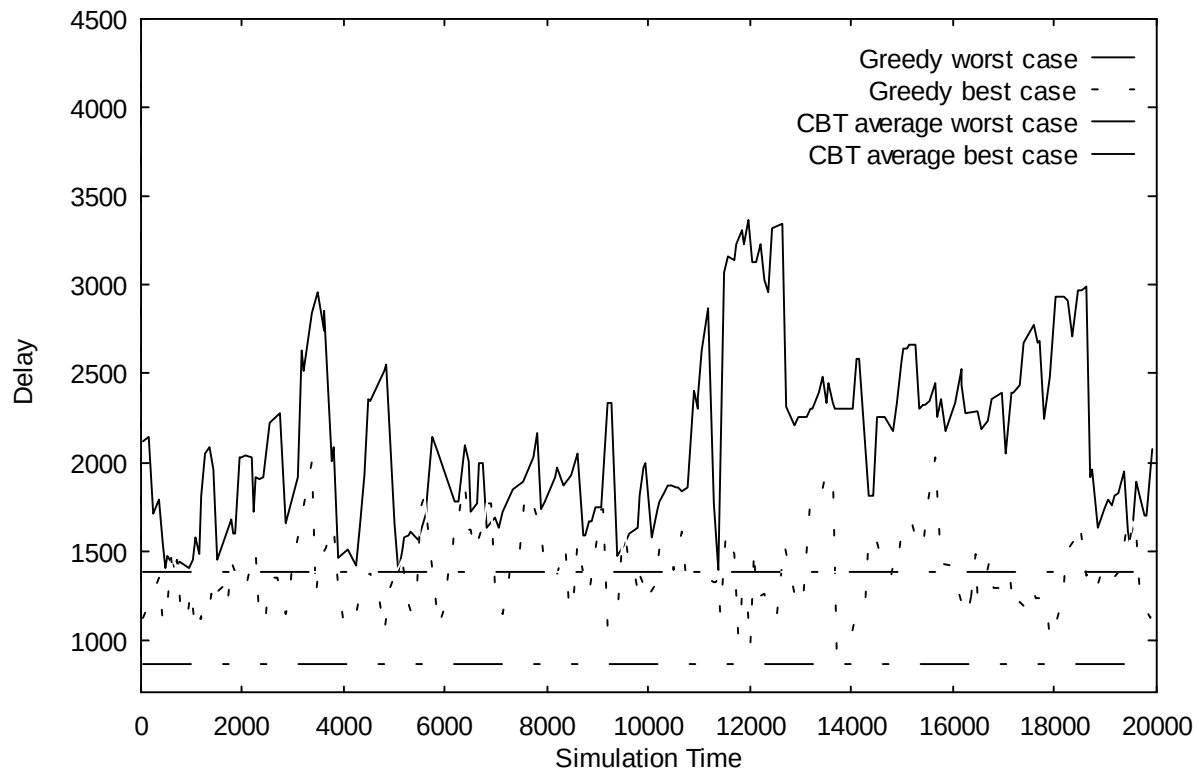

**Figure 8 Maximum delay pattern of GRD (single source, degree 5)**

CBT and SOPT show a similarly stable pattern over time. For CBT the gap between the worst case and the best case increases as more sources share the tree, whereas the gap for SOPT only increases marginally. The performance patterns of the other 3 algorithms: GRD, TOPT and MDT showed more severe fluctuations. However the fluctuation is stabilised as more static sources share the tree. Generally, the initial configuration of the tree is important in determining the shape and performance of the greedy type trees. Once the main stem of a tree route is in place, subsequent modification of branch paths only causes minor changes to the overall performance. This explains why the patterns of greedy type trees stabilise when many sources share the tree. As long as the source locations are static throughout the connection, the main stem of the tree is fixed and the performance is stable.

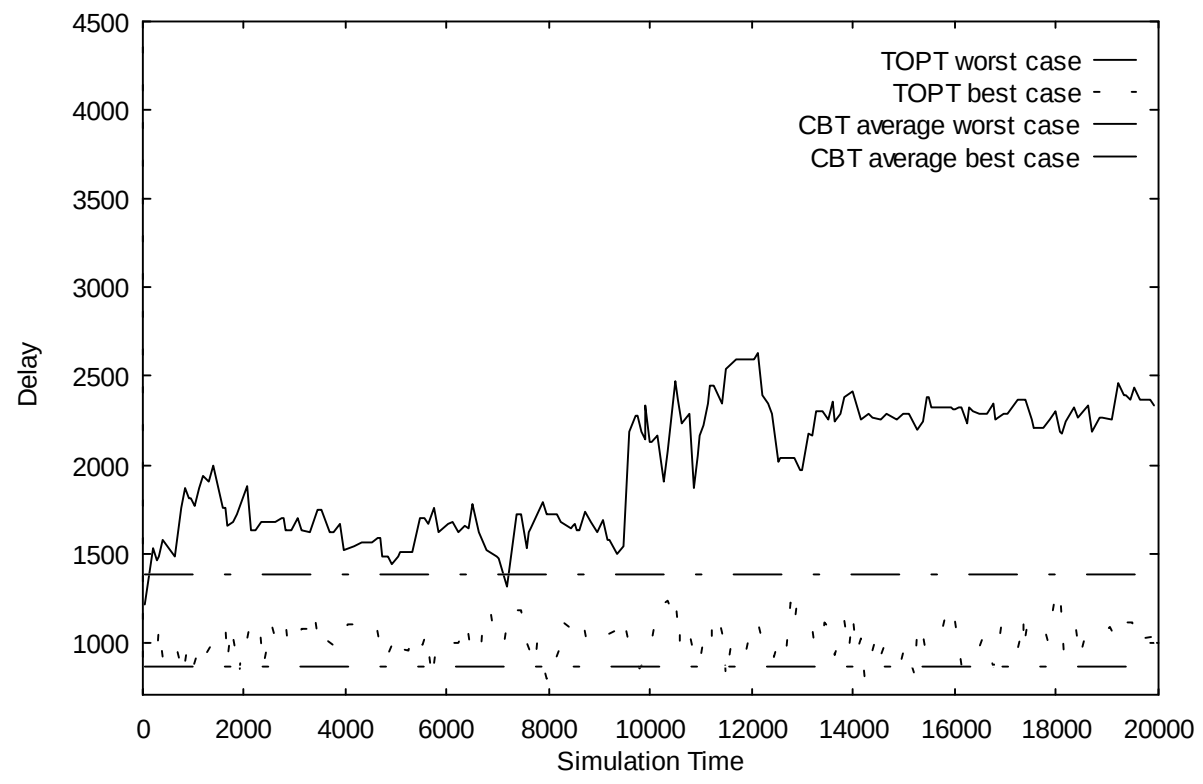

**Figure 9 Maximum delay pattern of TOPT (single source, degree 5)**

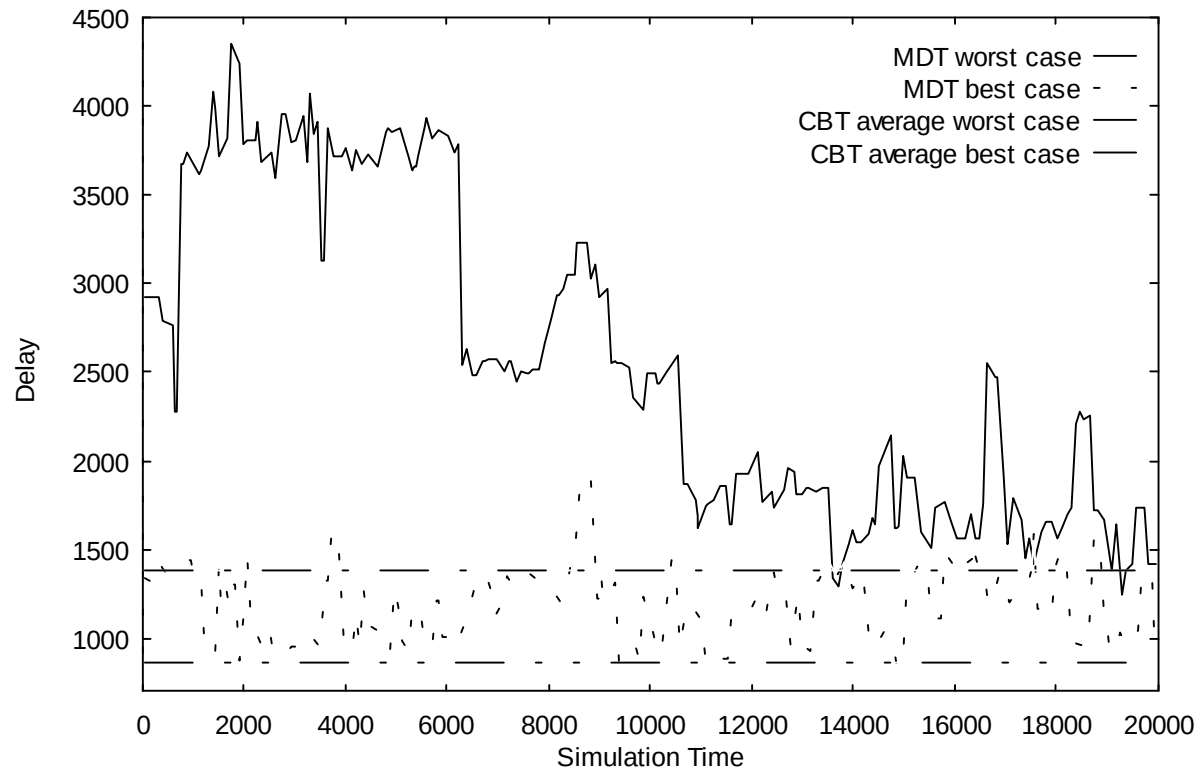

**Figure 10 Maximum delay pattern of MDT (single source, degree 5)**

Two horizontal lines are drawn in Figures 8~10 for comparison with the CBT average worst cases and average best cases. The worst case performance of TOPT and MDT stays below than CBT when multiple sources share the tree.

## 4.3 Effect of group size

Another factor in the performance of multicast trees is the efficiency for different numbers of sources and receivers. Figures 11~12 illustrate the changes in the average delay and number of links as the number of members in the group increases. The graphs presented are for cases where 5 sources share the tree in networks of degree 5. Other cases for different number of sources and different network degree showed a similar pattern, and only differed in the scale of the curves. As can be seen from the graphs, the delay in most of the algorithms grows slowly with increasing receivers, and is relatively stable above 30~40 receivers. Note that the total number of nodes in network is 200. Only GRD exhibits steady increase. There is little difference between the algorithms for link usage although GRD is always best and CBT is always worst. Generally the performance curves of delay by number of members fits the graph of the exponential function;

$$\mathbf{y = \mathit{h} - \mathit{a} * exp( -\mathit{x} / \mathit{b})}$$ , where ***h, a, b*** are the characteristic constants

More rigorous mathematical analysis is left for the future study issues.

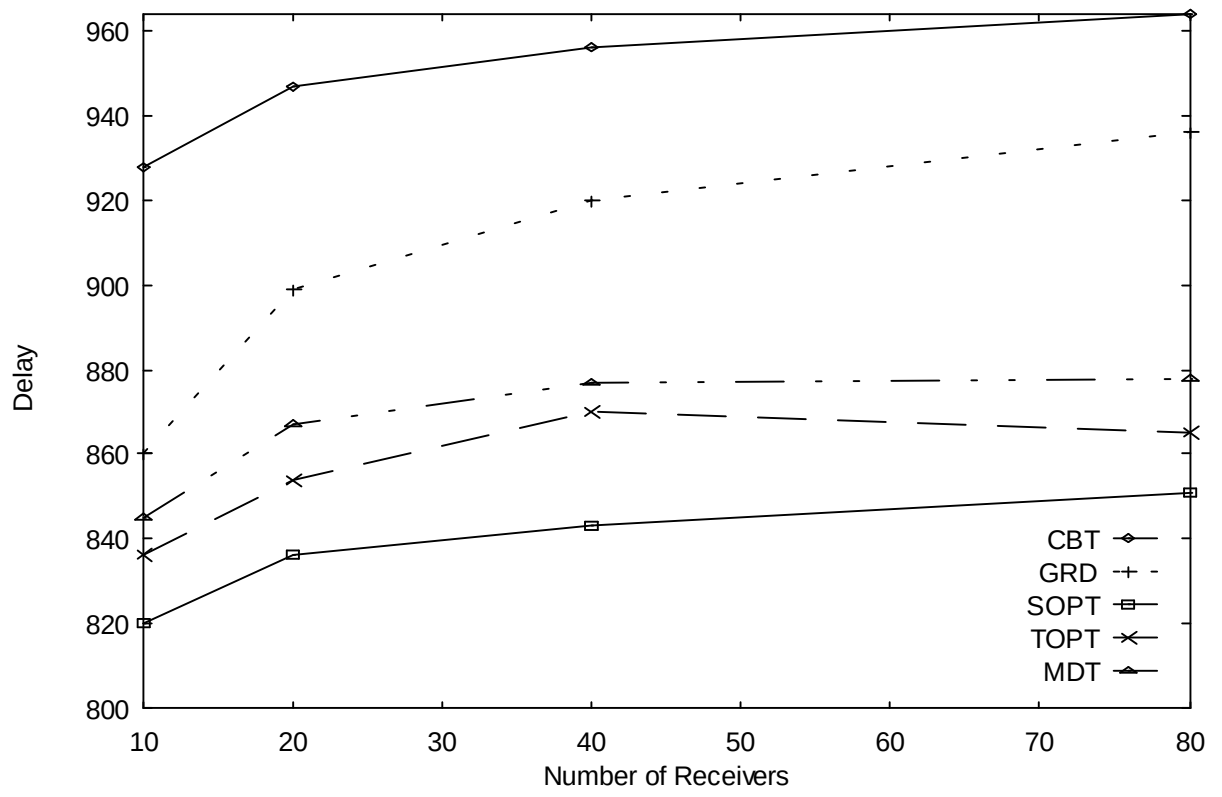

**Figure 11 Average delay by group size (5 sources in degree 5)**

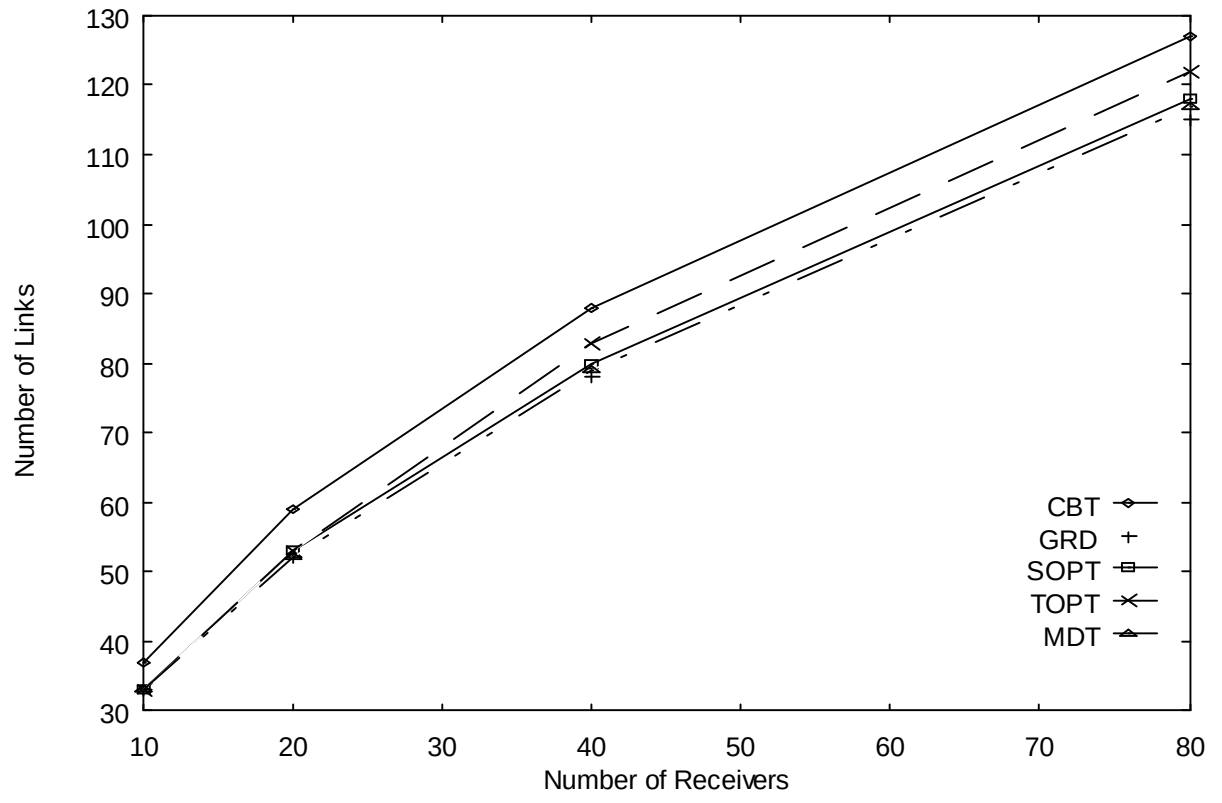

**Figure 12 Link usage by group size**

## 4.4 Effect of network degree

It is expected that the performance measures will display an improvement as the degree of the network increases. This is confirmed in this study as shown in Figures 13~14. This is a typical result with multicast trees, as has been reported elsewhere [2] [5] [7]. It is of note that the performance curves for the algorithms run in parallel, and the gap between them is consistent as the degree of connectivity increases. This means that an increase of network degree does not change the relative performance of the algorithms.

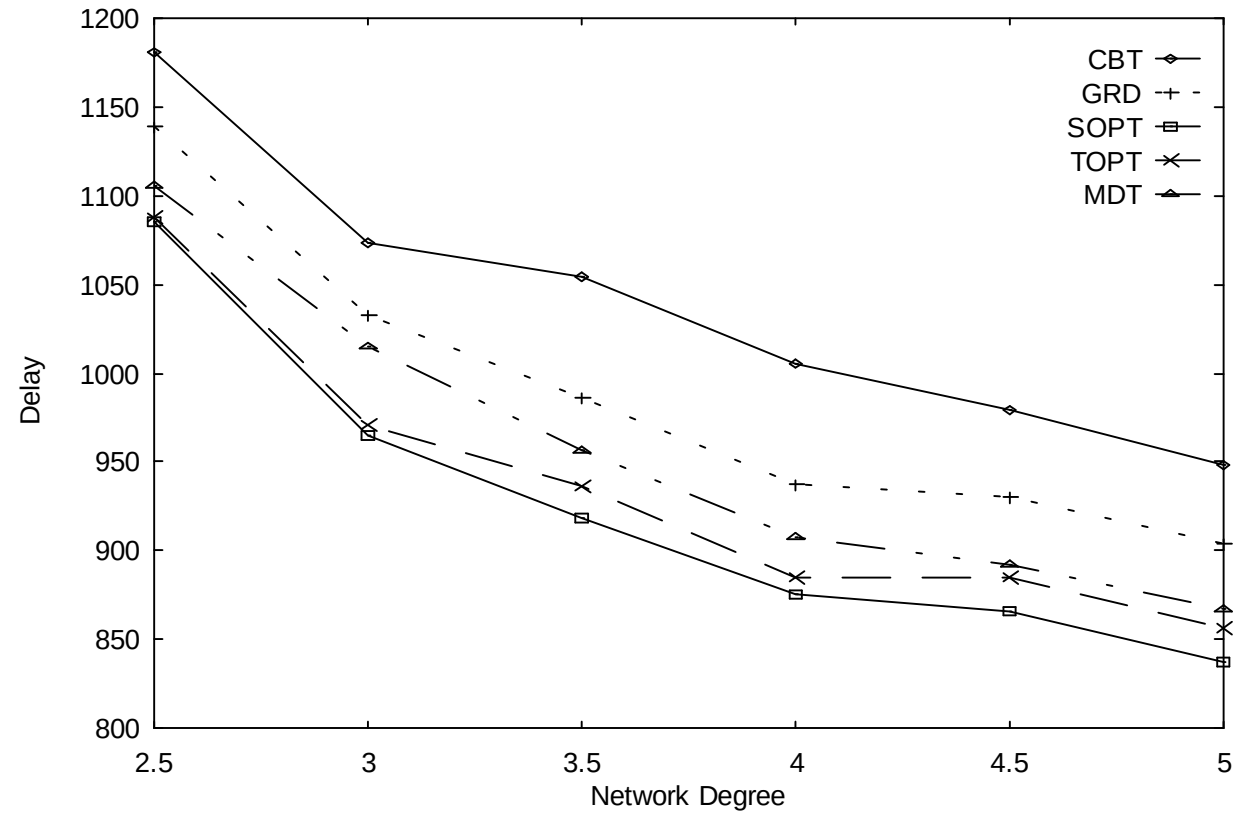

**Figure 13 Average delay by network degree**

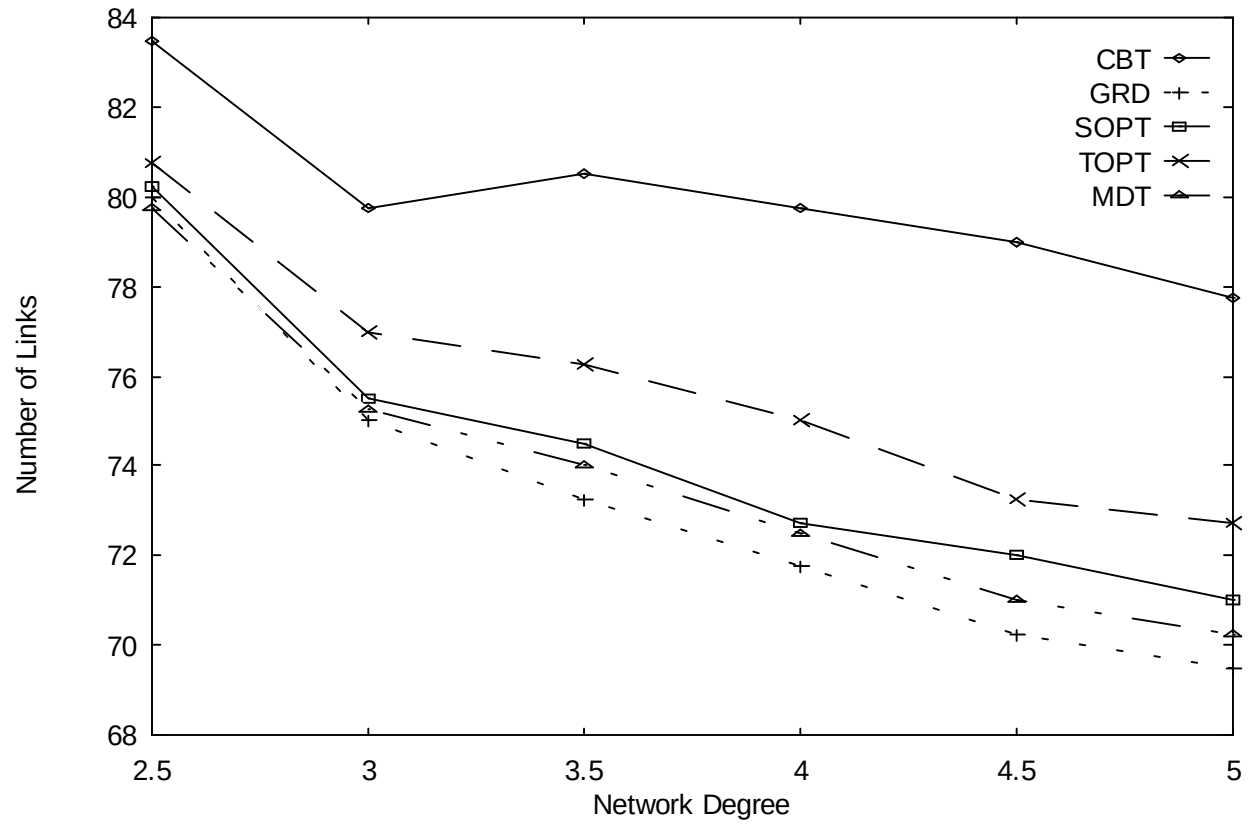

**Figure 14 Link usage by network degree**

## 4.5 Effect of multiple, static sources (fixed location)

Figures 15~16 compare the performance for an increasing number of static sources transmitting to variable number of receivers. The delay performance of CBT is best when the core node becomes the unique source node. However as many sources share the tree, the delay of CBT increases quickly and becomes worse than all the other algorithms except GRD. As was discussed in section 2, GRD does not constrain the maximum length.

SOPT is as good as CBT in single source multicast and it maintains a low delay as the number of sources increase. Delay of CBT is good when the core node becomes the only source node. In the presence of multiple sources, SOPT is significantly better than CBT. In general, SOPT exhibits up to 15% lower delay than CBT.

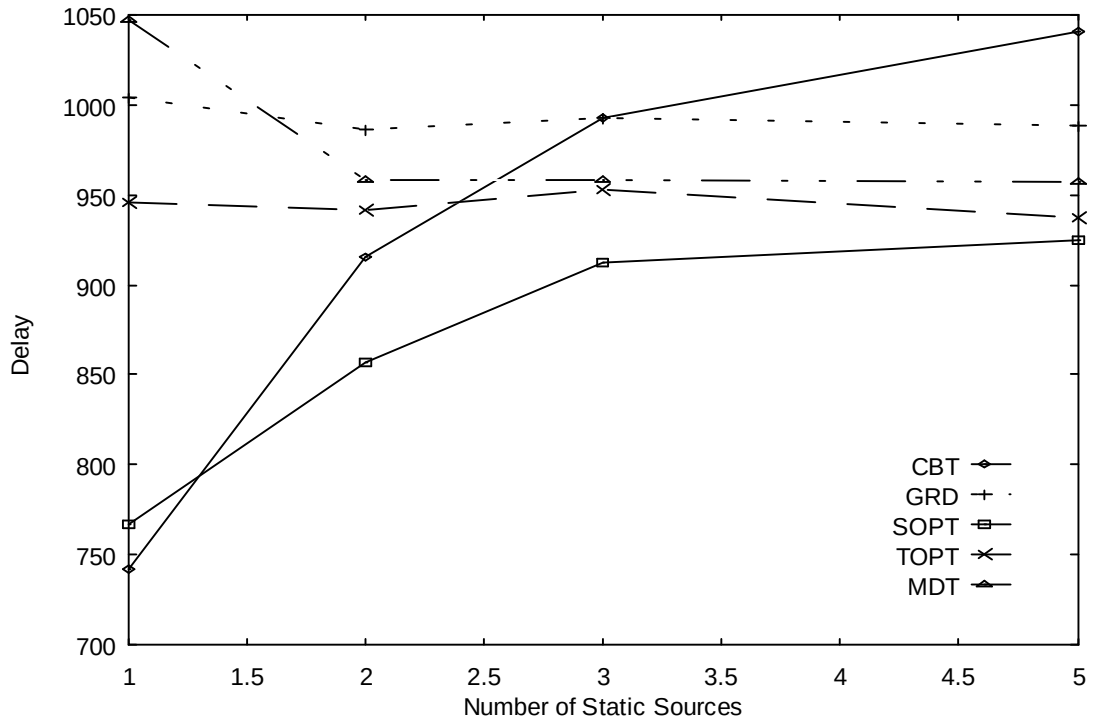

**Figure 15 Average delay by number of sources (fixed sources)**

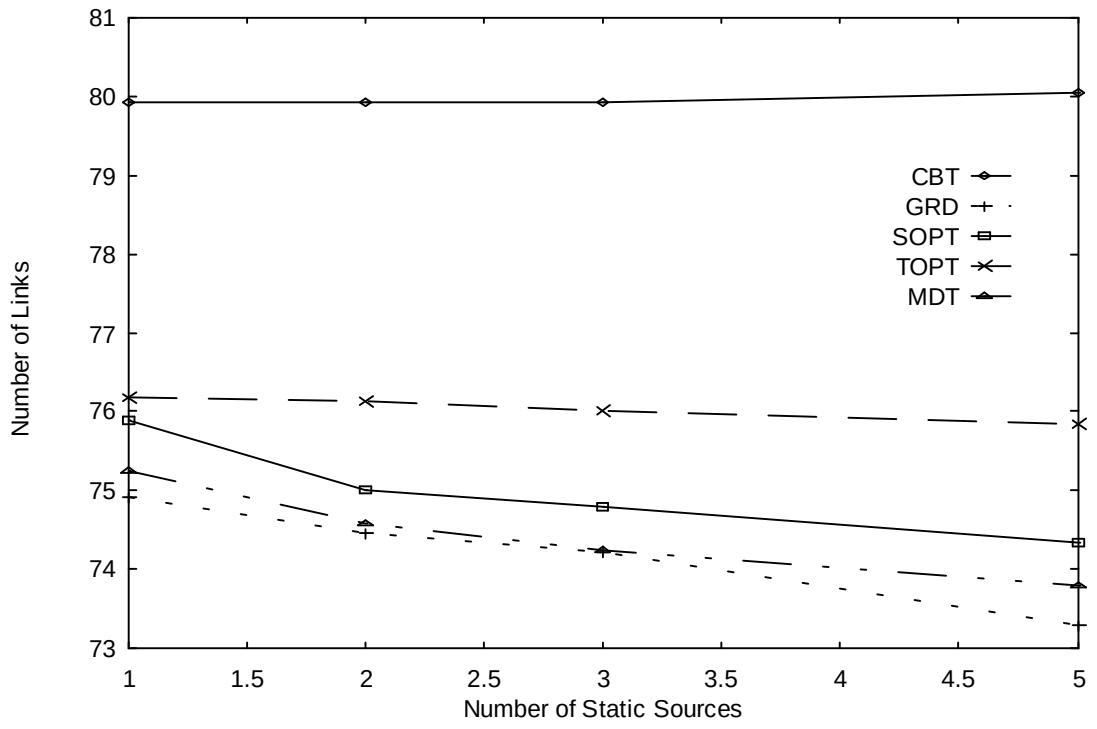

**Figure 16 Link usage by number of sources**

## 4.6 Effect of multiple sources (dynamic participation)

Figure 17 is the average delay performance graph of the dynamic source model, where the sources as well as receivers are exhibiting dynamic group membership, and a high proportion of nodes are sources. The pattern of the curves is basically same as the graphs of fixed sources, except that TOPT and MDT show lower delay than other algorithms. In particular, TOPT performs slightly better than SOPT when the proportion of the sources increases. This means that those two algorithms are relatively stable and efficient in applications where the source nodes are frequently changing.

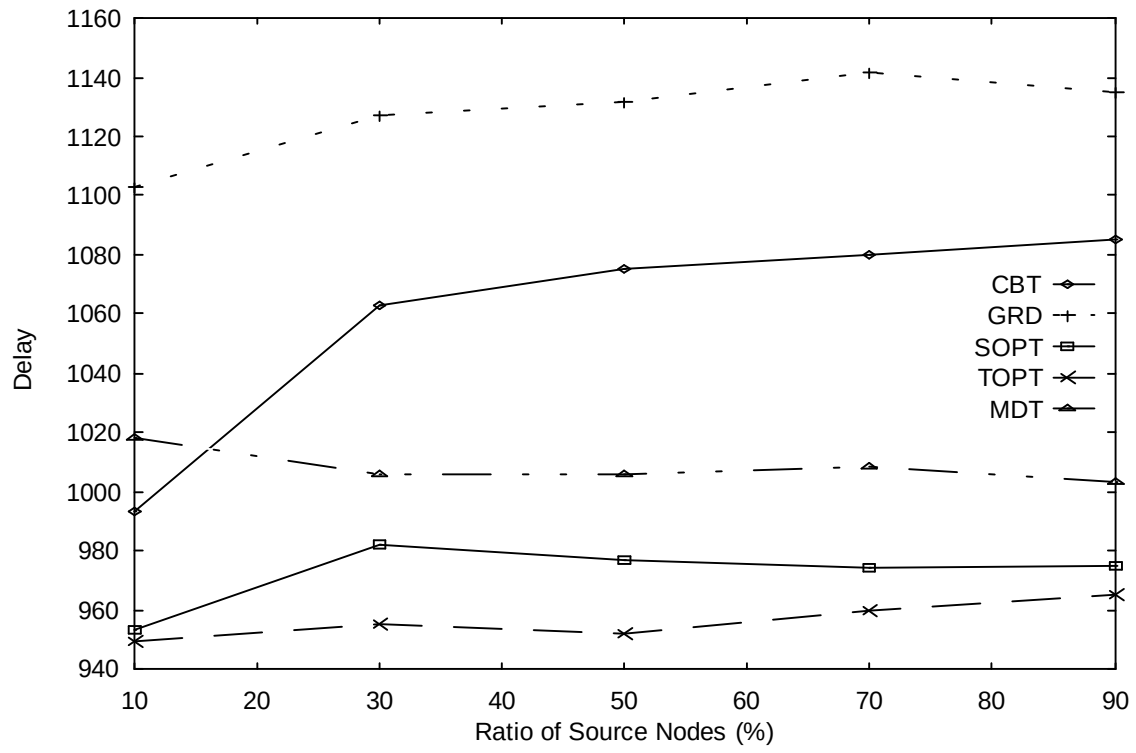

**Figure 17 Average delay by number of sources (variable sources)**

## 5. Conclusion

The delay and link usage performance characteristics of several new greedy based improvement algorithms are presented and discussed in this paper. More comprehensive simulation results can be found in [22] It is considered that SOPT is an efficient, dynamic, shared multicast tree in terms of delay and link usage. TOPT performs well when the participation of source nodes is dynamic, and in particular, builds well-balanced, efficient distribution trees when the members of group are densely populated. MDT did not show a satisfactory delay performance. Note that MDT does not construct a minimum diameter tree. (The solution to the problem of constructing a minimum diameter tree has been proven to be NP-Complete by Wall. [4]) Instead, MDT attempts to reduce the longest distance of the tree, and hence MDT may be worthwhile as a heuristic algorithm for producing an approximation to the minimum diameter tree.

On the whole, the improved greedy algorithms perform better than CBT in terms of delay and link usage. The algorithms are a new approach to share based multicast routing, and can be further improved to accommodate other performance criteria such as maximum throughput, low nodal overhead and low concentration of connections.

The greedy type algorithms may require more computational overhead and protocol complexity than shortest path algorithms. Modern flood routing methods [11] [12] enable the group information to be broadcast with low overhead. The algorithms proposed and evaluated in this paper have been developed primarily in the context of finding an efficient multicast algorithm utilising the specific features of flood routing. The development of multicast flood routing algorithm and protocol is part of on-going project at Monash University. Flood routing method for point-to-point connection has successfully been developed and implemented on the experimental Caroline ATM LAN network [10]. The design of a multicast flood routing protocol based on the SOPT algorithms is in progress.